\documentclass[aps,prl,twocolumn,superscriptaddress,preprintnumbers,amsmath
,amssymb,showpacs]{revtex4}

\usepackage{graphicx,amssymb,amsmath}

\begin{document}


\title{Nucleation and Growth of bundles of Single-Wall Carbon Nanotubes
 (C-SWNTs): the B\'enard-Marangoni Instability (BMI)  model}


\author{F.Larouche}
\email{larouche@inrs-emt.uquebec.ca}
\affiliation{INRS, \'Energie, Mat\'eriaux et T\'el\'ecommunications, Canada}
\author{J.Duquette}
\email{jonathan.duquette@mail.mcgill.ca}
\affiliation{Mechanical Engineering Department, McGill University, Canada}
\author{L.Cortelezzi}
\email{luca.cortelezzi@mcgill.ca}
\affiliation{Mechanical Engineering Department, McGill University, Canada}
\author{N.Nigam}
\affiliation{Mathematics and Statistics Department, McGill University, Canada.}
\author{B.Stansfield}
\affiliation{INRS, \'Energie, Mat\'eriaux et T\'el\'ecommunications, Canada}

\author{}
\affiliation{}


\date{\today}

\begin{abstract}
A complete explanation of the synthesis of metal-catalyst nanoparticles,
 and the subsequent nucleation and growth of bundles of C-SWNTs is
 introduced using a novel model. It is shown that the  synthesis process leads to the 
formation of a liquid layer
 supersaturated in carbon surrounding each metallic-catalyst nanoparticle. The onset of a solutal
 B\'enard-Marangoni instability and the subsequent formation of patterns
 of hexagonal convection cells in the liquid layer is predicted and
 quantified by linear and weakly nonlinear analyses. The nucleation
 and growth of a C-SWNT at the center of convection cell is explained.
\end{abstract}


\maketitle
This Letter presents a model which describes the mechanism for the 
 nucleation and growth of carbon  single-wall nanotubes (C-SWNTs). Several
 models have been proposed in the literature \cite{Saito,Gorbunov2,Micro,
Nouveau2};  they have succeeded in describing the general scenario
 leading to growth of C-SWNTs. Our model, however, is able to explain 
specific and important aspects such as the nucleation, the individual
 structure of C-SWNTs and their diameter. In brief, our model explains how:
 1. the synthesis conditions for C-SWNTs lead to the formation of
 carbide-metal nanoparticles surrounded by a layer of liquid supersaturated 
in carbon; 2.  a B\'enard-Marangoni instability is generated in this
 liquid layer; 3. the B\'enard-Marangoni instability leads to the
 formation of hexagonal convection cells, at the center of which
 nanotubes nucleate; 4. the C-SWNTs grow by a root-growth mechanism at least initially and 5. the BMI offers a simple kinetic explanation for the diameter of the C-SWNTs.

 Recently, a review of the nucleation and growth of C-SWNTs \cite{frederic}
 suggested that an interface instability is responsible for the nucleation
 of C-SWNTs. Leveraging this idea, we develop a model based on the
 B\'enard-Marangoni instability (BMI), called the BMI model. In this 
Letter, we show that nucleation of C-SWNTs indeed results from the
 onset of the BMI. For concreteness, we study the common case of a 
long bundle growing on one side of a large (5nm to 20nm) nanoparticle
 (NP) and only consider an iron catalyst, which is known to result
 in synthesis of C-SWNTs  \cite{Smiljanic}. The treatment remains
 valid for any catalyst.
 
\begin{figure}
\includegraphics[width=0.5\textwidth]{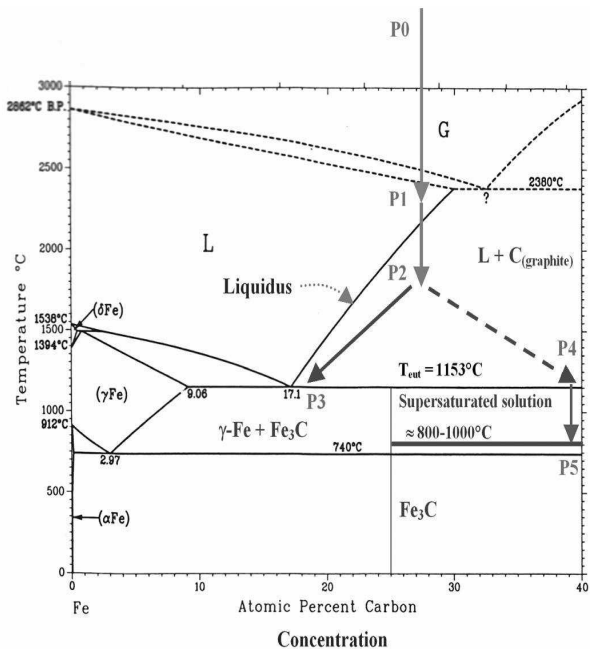}
\caption{\label{Fig1}The overall scenario for growth of C-SWNTs
 represented in the Fe-C phase diagram}
\end{figure}

We first describe the formation of the nanoparticle and the liquid
 layer surrounding it,  by analysing the iron and carbon (Fe-C) phase
 diagram (Fig.\ref{Fig1}). In gas-phase processes, the synthesis begins
 by the creation of a mixture of Fe-C vapors in an inert gas. The
 Fe-C mixture is initially at a high temperature ($5000^{\circ}C$)
 (point {P0} in Fig.\ref{Fig1}) and contains a fraction of about
 2\%-3\% of iron atoms with respect to carbon atoms. The vapor
 mixture is then cooled very rapidly ($\approx 10^{6}$ $^{\circ}C/s)$
. As the temperature decreases to around $2300^{\circ}C$ ({P0 to P1
 in Fig.1}), the vapor mixture condenses to form nanometric droplets
 of an iron-carbon solution having a concentration of about 25\%-30\%at.
 If the cooling rate remains high, the iron-carbon droplets  undergo
 a non-equilibrium solidification process down to around $1500^{\circ}C$,
 maintaining droplets in a supersaturated state
 ({P1 to P2 in Fig.\ref{Fig1}}). Subsequently, droplets experience
 a rapid segregation process \cite{Micro} so as to recover their
 equilibrium state. At this point,  the carbon concentration in the
 droplet is higher than the equilibrium liquidus value. As a
 consequence, carbon is expelled radially toward the surface of 
the droplet, until the composition of the droplet center reaches 
the liquidus equilibrium. From this point, the core of the droplet
 follows a different path in the phase diagram than the  
layer surrounding this core. 

We now discuss the evolution of the droplet core. This core solidifies 
at the eutectic temperature, $T_{eut}$ ({P2 to P3 in Fig.\ref{Fig1}}),
 by  a eutectic transformation that results in the formation of a carbide 
($Fe_{3}C$) phase and of a $\gamma-Fe$ phase. In agreement with this
 analysis, the presence of a thin carbide layer has already been observed 
\cite{Saito, Nouveau2} by transmission electron microscopy at the 
surface of the core. The analysis is further supported by  Saito \cite
{Saito} and Alvarez {\it et al.} \cite{Alvarez}, who showed that the
 synthesis temperature of C-SWNTs and their growth are correlated to
 the catalyst-carbon eutectic temperature.

In the outer layer of the droplet, the expelled
 carbon will not have enough time to nucleate into small graphitic
 flakes. Carbon remains dissolved in atomic form in the liquid metal,
 resulting in the formation of a nanometric liquid layer surrounding
 the solid core. This liquid layer is supersaturated in carbon \cite{Russe}, with a 
concentration of as much as  40\%-50\%at.({P2 to P4 in Fig.\ref{Fig1}}). We note that the liquid phase
 still exists few hundreds of degrees below $T_{eut}$ \cite{Russe, Russe3}
 since a solution having a non-equilibrium carbon concentration
 ($>25\%$at. for Fe-C, Ni-C and Co-C binary systems) has a lower 
melting temperature than the eutectic temperature of the system
 at equilibrium. We can thus define a non-equilibrium eutectic
 temperature for the supersaturated solution. The melting point 
of this solution can vary from $800^{\circ} C$ to $1000^{\circ}
 C$ and will depend on the carbon concentration, a higher
 concentration leading to a lower melting point. Therefore, the
 surface liquid layer can exist from {P4 to P5} in the phase
 diagram Fig.\ref{Fig1}. This is also supported by Gorbunov {\it et al.}
 \cite{Gorbunov2} who proposed that the catalytic nanoparticles 
involved in the C-SWNTs synthesis are molten.

We now focus on a single NP, and explain a mechanism for the nucleation
 and growth of nanotubes, driven by a convective instability. A NP is
 composed of a liquid layer supersaturated in carbon surrounding a solid 
core composed of a carbide phase and of a $\gamma-Fe$ phase saturated in 
carbon. We assume that after the segregation process responsible for the
 generation of the liquid layer, the concentration in this layer will vary
 from 25\%at. at the solid-liquid interface to 45\%at. at the liquid-gas 
interface.  We claim that the  B\'enard-Marangoni convection governs the
 carbon transport in the liquid layer. The instability can be driven by
 temperature or concentration gradients; our calculations will demonstrate 
that concentration effects dominate.

Since C-SWNTs grow in a bundle from one side of the NP, we neglect
 curvature effects, and  work with  a planar geometry. We consider a
 liquid layer of thickness $d$ and of infinite horizontal extent, with the same carbon concentration gradient as above.
 The free surface will be considered to be 
 undeformable. Gravity effects can be neglected because Rayleigh-B\'enard
 convection is negligible compared to the B\'enard-Marangoni convection,
 as will be shown by our calculation. We model the Fe-C solution as 
a Newtonian and incompressible fluid.

Since the BMI is driven  by variations in surface tension $\sigma$,
 which in turn are caused by gradients in concentration $\tilde{c}$,
 it is crucial to establish a mathematical relationship between 
these two quantities. Unfortunately, there is no experimental data 
available for such non-equilibrium systems. Consequently, we extend 
the linear dependence $\sigma=\sigma_0+\gamma_c(\tilde{c}-c_0)$
 \cite{Bragard}, valid in the eutectic region,  to the entire
 hyper-eutectic region ($\tilde{c} > 17.1\% at$.), where carbon acts
 as a surface tension increasing solute. Here $\sigma_0$ and $c_0$
 are the eutectic composition surface tension and concentration,
 respectively. The solutal surface tension coefficient, $\gamma_c$,
 is assumed constant. We assume the liquid layer is initially in an
 unperturbed state.  The fluid is at rest and the initial concentration 
profile $\overline{c}$ is considered linear with respect to the vertical
 distance $Z$, $\overline{c}=c_{SL}+\beta Z$, where $\beta$ is a (positive)
 constant concentration gradient and $c_{SL}$ is the carbon concentration
 at the solid-liquid interface.

 The motion of the liquid layer is described by the non-dimensionalized
 continuity and  
Navier-Stokes equations:
\begin{equation} \nabla \cdot \vec{u}=0, \qquad 
Sc^{-1}\left( \partial_{t}\vec{u}+\vec{u}\cdot\nabla\vec{u}\right)=-\nabla 
p + \nabla^{2}\vec{u},
\end{equation}
where $(x,y,z)$ are the three Cartesian coordinates, $\vec{u}=(u,v,w)$ is the velocity, $p$ the pressure, 
$Sc=\mu/\rho D_{L}$ the Schmidt number, $\mu$ the viscosity, 
$\rho$ the iron-carbon solution density and $D_{L}$ the carbon
 diffusivity in the liquid phase. Here, we have scaled distance,
 time, velocity, concentration and pressure  by $d$, $d^2/D_{L}$,
 $D_{L}/d$, $\beta d$ and $\mu D_{L}/d^{2}$, respectively.
We couple the above equations to a solutal diffusion process in the 
liquid layer 
modeled by:
\begin{equation}
\partial_{t}c+\vec{u}\cdot\nabla c=\nabla^{2}c - w,
\end{equation}
where $c=\tilde{c}-\overline{c}$ is the deviation from the unperturbed-state
 concentration $\overline{c}$. 

At the solid-liquid interface $(z=0)$, the velocity is zero, $\vec u=0$,
 there is no flux of carbon, $\partial_{z}c\vert_{z=0}=0$, and the carbon concentration is $c_{SL} =25\%at$. At the liquid-gas
 interface $(z=1)$, the surface is undeformable, $w\vert_{z=1}=0$,
 and the initial surface concentration of carbon is $c_{LG} = 45\%at.
 $ Depending on the mass transport generated by the solutal BMI, $c_{LG}$ 
will change with time.

At the onset of the instability, the boundary condition expressing the
 diffusion of carbon across the surface is given by $
\partial_{z}c\vert_{z=1} =0. $
As the liquid layer is supersaturated with carbon at the onset of the 
instability, neither absorption nor adsorption are possible. Thus, the 
solutal Biot number {\it Bi} (characterizing rate of absorption) and the 
adsorption number {\it Na} \cite{Brian} are both zero.

At the liquid-gas interface, the tangential stress is determined by the 
surface tension and takes the following nondimensional form:
\begin{equation}
\partial_{zz} w=-Ma_{s}(\partial_{xx} c +\partial_{yy} c),
\end{equation}
where the dimensionless solutal Marangoni number 
\begin{equation}
Ma_{s}=\frac{\gamma_{c} d(c_{LG}-c_{LS} )}{\mu D_{L}},
\end{equation}
is an important predictor of the onset of the instability. The critical 
value of the solutal Marangoni number, $Ma^{c}_{s}$, above 
which the instability is generated in the case where {\it Na} and {\it 
Bi} are zero, is found from previous linear stability analyses
 \cite{Brian, Bragard} to be $Ma^{c}_{s}=50$. We will now show that
 $Ma_{s}>Ma^{c}_{s}$ under the particular synthesis conditions of C-SWNTs.

 We know from experimental observations that the diameter of the core of
 a NP is generally between 5nm and 20nm \cite{Nouveau2}. We postulate
 that the typical thickness of the liquid layer surrounding the core is
 between 2nm and 4nm. The value of the solutal surface tension coefficient
 $\gamma_{c}$ has been determined in the literature \cite{Metal} to be
 0.03 $N/m\cdot\%at.$ We assume that the carbon diffusivity $D_{L}$ in a
 highly supersaturated solution is equivalent to substitutional diffusion
 because interstitial sites are mostly filled. In this case, the diffusivity
 is almost independent of the nature of the solute for liquid metals and
 varies between $10^{-9} m^{2}/s$ and $10^{-8} m^{2}/s$ \cite{Materiaux}.
 We choose the lower value of diffusion since the liquid layer is
 supersaturated. Finally, we use a viscosity within the range $\mu=8 \pm 2 mPa\cdot s$, based on  different extrapolations from data 
\cite{Steelmaking}. The solutal Marangoni number describing our system 
is then estimated to be between  $Ma_{s}=120$ to $400$, which is above
 the critical Marangoni number $Ma^{c}_{s}$ required to generate the
 instability. In contrast, the thermal and solutal Rayleigh numbers and 
the Marangoni thermocapillary number are estimated to be of the order of $10^{-20}, 10^{-11}
 \text{ and } 10^{-3}$ respectively, far below their respective critical
 values. It follows from these calculations that the solutal Marangoni
 effect dominates all the other effects.

Having described the onset of the B\'enard-Marangoni instability, we now
 determine the shape of the convection cells. For this, we rely on the
 results of the weakly nonlinear analysis done by Bragard {\it et al.} The
 parameter necessary to determine the pattern is the distance from the
 threshold of the instability \cite{Bragard}:
\begin{equation}
\epsilon=\frac{Ma_{s}-Ma^{c}_{s}}{Ma^{c}_{s}}. 
\end{equation}
The value of $\epsilon$ varies between $1.4$ and $7$ depending on the thickness 
of the liquid layer, and can only be considered to be approximate,
 given the imprecision in the experimental parameters. Using a linear
 extrapolation (Fig.11 and Table 1 from \cite{Bragard}), we determined
 the following stable configurations for the convection cells when 
Na = Bi = 0: 
\begin{center}
\begin{tabular}{|c|c|}
\hline
Stable configurations & $\epsilon$ \\
\hline \hline
Hexagons & $0 < \epsilon < 2.4$ \\
\hline
Hexagons, rolls & $2.4 < \epsilon < 6.3$ \\
\hline
Rolls, hybrid cells & $\epsilon >  6.3$ \\
\hline
\end{tabular}
\end{center}

We conclude that the hexagonal stable configuration is favoured in the major 
part of the range of $\epsilon$. Note that we expect rolls and hybrid cells 
to be unfavorable for the nucleation of C-SWNTs.

\begin{figure*}
\includegraphics[width=1.0\textwidth]{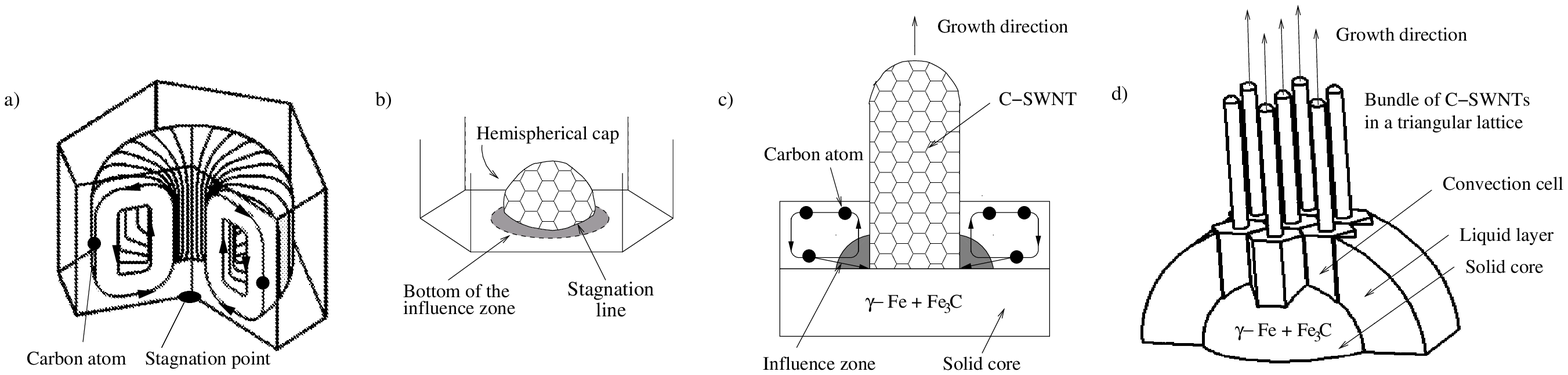}
\caption{\label{Fig3}Steps of the growth: a) Flow in one hexagonal convection
 cell, b) Formation of a C-SWNT cap on the solid-liquid interface, c) Growth
 of a C-SWNT in one hexagonal convection cell, d) Collective growth:
 a bundle of C-SWNTs}
\end{figure*}

We now focus on the nucleation and growth of a C-SWNT at the center of a
 hexagonal convection cell. There is no mass transport between cells, and
 therefore it suffices to consider one convection cell. We know from fluid
 dynamics that in each convection cell, the liquid rises in the center of
 the hexagon and descends along the edges of the hexagon, forming a vortex
 ring \cite{Chandra} (Fig. \ref{Fig3}a). The center of the top face and
 bottom face of a cell are stagnation points, where the velocity of the
 fluid is zero. The stagnation point at the bottom of a cell, lying on
 the solid-liquid interface, is of crucial importance since we postulate
 that a C-SWNT  nucleates at this point. The tube grows by  incorporating
 carbon atoms from the liquid.

We envision the beginning of the nucleation process as follows: a fraction
 of the carbon atoms are transported close to the stagnation point on the
 carbide layer  by the convective fluid flow (Fig.\ref{Fig3}a). The velocity 
of carbon atoms decreases close to the stagnation point, reducing their
 kinetic energy to below the adhesion energy on the metal catalyst. There is 
thus a semi-spherical influence zone around the stagnation point in which
 carbon atoms move slow enough to be adsorbed onto the catalyst. The 
semi-spherical shape is imposed by the axial-symmetry of the flow near the
 center of the cell. This influence zone acts as a heterogeneous nucleation 
site. Once the carbon is adsorbed at the solid-liquid interface within this
 zone, it crystallizes under its minimum energy configuration. It has been 
shown \cite{Fan} that the nucleation of a closed hemispherical cap is 
favoured in presence of a metal surface.

As the cap is formed, the topography of the bottom of the cell changes. The
 stagnation point is lost while a stagnation line (circumference) at the base
 of the cap is created (Fig. \ref{Fig3}b). The shape of the influence zone
 also changes: it becomes a sector of a torus surrounding the stagnation line
 (Fig. \ref{Fig3}b-c). The hemispherical shape of the cap, observed in 
experiments \cite{Micro}, guarantees that the angle of contact between the
 cap and the carbide layer is nearly vertical. Consequently, the carbon atoms
 decelerate sufficiently and can be incorporated between the cap and the 
solid surface. The binding of the cap with the metal surface prevents its 
closure at the bottom and allows the root-growth \cite{Micro} of a
 cylindrical structure, the C-SWNT (Fig. \ref{Fig3}c). We postulate that the
 carbon supply at the base of the cap, which is enhanced by the convective
 flow, exceeds the rate at which bonds in the cap can rearrange. This
 prevents the lateral growth of the cap and allows the C-SWNT extrusion with
 a constant diameter \cite{Fan}. Since the size of a convection cell is
 known to be of the order of the liquid layer thickness \cite{Chandra}, we
 can estimate the diameter of the C-SWNT to be a few nanometers. This agrees with the diameter of experimentally observed  C-SWNTs, (0.7nm-3nm), \cite{Micro}. Recent total energy calculations using density functional theory predict a larger diameter for the nanotubes,and conjecture a kinetic factor limiting this diameter, \cite{Fan}. Our BMI model provides precisely such a kinetic explanation.  Note that outside the influence zone, it is
 more favourable for carbon atoms to follow the convection flow than to
 bind laterally on the C-SWNT since graphitization can only occur
 at much higher temperatures.

The collection of hexagonal convection cells is responsible for the triangular
 lattice structure of the bundles of C-SWNTs (FIG. \ref{Fig3}d) observed in
 experiments at the onset of the growth \cite{Micro}. After the C-SWNTS grow beyond a certain point their organization may be influenced by other factors, but this stage of the growth is outside the scope of our model.
Our model explains the nucleation and the beginning of the growth of C-SWNTs 
in the absence of absorption ($Bi=0$). However, the carbon atoms initially 
contained in the liquid layer are not sufficient for the growth of long
 bundles of C-SWNTs. Carbon atoms must therefore be absorbed from 
the surrounding gas to maintain the growth process. The extension of the present
 model to the case when $Bi\ne 0$ is not trivial and will be addressed in a
 future study.

In summary, using the Fe-C phase diagram, we have explained why a high
 cooling rate is necessary in the synthesis process to generate a layer of 
liquid metal surrounding the NP. Leveraging a linear stability analysis, we
 have shown that the solutal BMI can be generated in the liquid layer.
 Exploiting a weakly nonlinear analysis, we have shown that the hexagonal
 pattern can be favoured under the synthesis conditions considered. Once
 initiated, the BMI is seen to be responsible for the nucleation and growth
 of C-SWNTs.

\end{document}